\documentclass{elsart}

\usepackage{graphicx,color}
\usepackage{amssymb}
\usepackage{amsmath}
\usepackage{latexsym}
\usepackage{amsxtra}
\usepackage{amscd}

\journal{Physics Letters B}

\begin{document}

\begin{frontmatter}

\title{First identification of large electric monopole strength in well-deformed rare earth nuclei}

\author[lmu]{K. Wimmer},
\author[tu]{V. Bildstein},
\author[tu]{K. Eppinger},
\author[tu]{R. Gernh\"auser},
\author[lmu]{D. Habs},
\author[tu]{Ch. Hinke},
\author[tu]{Th. Kr\"oll},
\author[tu]{R. Kr\"ucken},
\author[lmu]{R. Lutter},
\author[lmu]{H.-J. Maier},
\author[tu]{P. Maierbeck},
\author[lmu]{Th. Morgan},
\author[lmu]{O. Schaile},
\author[lmu]{W. Schwerdtfeger},
\author[tu]{S. Schwertel} and
\author[lmu]{P.G. Thirolf}

\address[lmu]{Fakult\"at f. Physik, Ludwig-Maximilians-Universit\"at M\"unchen, 85748 Garching Germany}
\address[tu]{Physik Department E12, Technische Universit\"at M\"unchen, 85748 Garching Germany}

\begin{abstract}
Excited states in the well-deformed rare earth isotopes $^{154}$Sm
and $^{166}$Er were populated via ``safe'' Coulomb excitation at the
Munich MLL Tandem accelerator. Conversion electrons were registered
in a cooled Si(Li) detector in conjunction with a magnetic transport
and filter system, the Mini-Orange spectrometer. For the first excited
$0^+$ state in $^{154}$Sm at 1099~keV a large value of the
monopole strength for the transition to the ground state of
$\rho^2(\text{E0};\, 0^+_2 \rightarrow 0^+_\text{g}) = 96(42)\cdot 10^{-3}$ 
could be extracted. This confirms the
interpretation of the lowest excited $0^+$ state in $^{154}$Sm as
the collective $\beta$-vibrational excitation of the ground state. 
In $^{166}$Er the
measured large electric monopole strength of $\rho^2(\text{E0};\,
0^+_4 \rightarrow 0^+_\text{g}) = 127(60)\cdot 10^{-3}$ clearly
identifies the $0_4^+$ state at 1934~keV to be the $\beta$-vibrational 
excitation of the ground state.
\end{abstract}

\end{frontmatter}

\section{Introduction}\label{intro} 
The structure of excited $0^+$ states in deformed
even-even nuclei is still a matter of controversial discussion
despite intensive investigation. Traditionally the first excited
$0_2^+$ state has been interpreted as the $\beta$-vibrational
excitation of the ground state. However, in many nuclei the $0_2^+$
state has only weak transitions to the ground-state band, while
strong electric quadrupole transitions to the $\gamma$ band have
been found \cite{casten94}. This contradicts the traditional
interpretation, since a transition from a $\beta$-vibrational state
to the $\gamma$ band is suppressed due to the destruction of a
$\beta$ phonon and, at the same time, the creation of a $\gamma$
phonon. In this picture a $\beta$-vibrational state is characterized
by a strong transition to the ground-state band, namely
by a large $B(\text{E2};\, 0_\beta^+\rightarrow 2_\text{g}^+)\approx 10$~W.u. 
value and a strong E0 transition to the ground state with $\rho^2(\text{E0})\approx 100\cdot10^{-3}$ \cite{garrett_beta}. Only in very few
cases, such as $^{154}$Sm \cite{kruecken} and $^{166}$Er
\cite{garrett}, it has been possible to identify candidates for a
$\beta$-vibrational state by $\gamma$ spectroscopy. The unclear
situation led to an intense debate about the structure of low-lying
$0^+$ states. Based on calculations using the interacting boson
approximation (IBA) \cite{arima75,arima78}, Casten and von Brentano
\cite{casten94} have proposed that the $0_2^+$ state in deformed
nuclei should be interpreted as a second $\gamma$ phonon excitation built 
on the $\gamma$ vibration. Since in many cases the excitation energy of the
$0_2^+$ state is located below the $\gamma$ band and $B(\text{E2})$
values to the ground-state band as well as to the $\gamma$ band show
large fluctuations, this interpretation has been challenged by Burke
and Sood \cite{burke95}, Kumar \cite{kumar} and G\"unther \cite{guenther}.

In the original work by Casten and von Brentano, it was assumed that
the deformed nuclei are best described by a small area in the
parameter space of the IBA, which led to the prediction of the
character of the $0_2^+$ state in deformed nuclei as a two phonon
$\gamma\gamma$ vibration. In the framework of the simplified ECQF 
formalism \cite{lipas} nuclei are described by two parameters, $\zeta$ 
and $\chi$ and two scaling factors for energies and transition rates, 
respectively.
Recent work by McCutchan et al. \cite{mccutchan} mapped the position 
of the deformed nuclei for different isotopic chains of rare earth 
nuclei within the IBA symmetry triangle, revealing that the IBA
parameters to describe the low-lying structure of these nuclei can
differ significantly. The position within the symmetry triangle for
well-deformed nuclei was later related to the underlying
single-particle structure near the Fermi surface and the resulting
quasi-particle structure of the $\gamma$-vibrational state
\cite{hinke}.

It was also shown in recent years that the IBA consistently predicts
that the E0 strength from the first or second excited $0^+$ state in
deformed nuclei is large \cite{brentano}. Near the $U(5)-SU(3)$ leg 
($\chi=-\sqrt{7}/2$) the
$0_2^+$ state carries the E0 strength, while near the $O(6)$ corner,
the $0_3^+$ state exhibits large E0 strength. In an area in between the 
strength is shared among the $0_2^+$ and $0_3^+$ states.
This IBA prediction for well-deformed nuclei is not confirmed
experimentally, due to the lack of measured $\rho^2(\text{E0})$ values of
the E0 strength for excited $0^+$ states in these nuclei. For the few measured
examples, such as $^{166}$Er and $^{172}$Yb, where a small E0 strength
was observed for the $0_2^+$ states, it is not clear if these $0^+$ 
states correspond to those for which the IBA predicts
large $\rho^2(\text{E0})$ values. It is therefore important to obtain more
experimental data on E0 strength in well-deformed nuclei, which may
also lead to new insights in the nature of the low-lying $0^+$
states.

It is the purpose of this letter to report on the first observation
of large E0 strength of excited $0^+$ states in the well-deformed
nuclei $^{154}$Sm and $^{166}$Er, which confirm the interpretation
of the $0_2^+$ and $0_4^+$ states, respectively, as $\beta$-vibrational
states. Before we describe the details of the
performed experiments and their results in the next sections, we
will briefly review the existing information on $0^+$ states in
$^{154}$Sm and $^{166}$Er. 
Due to the existence of the systematic parameter studies \cite{mccutchan} 
and the
explicit predictions of the E0 strength in well-deformed nuclei \cite{brentano}
within the framework of the IBA, we will concentrate our discussion in
the final section on a comparison with IBA calculations. Although a similar
comparison could and should be done on the basis of collective models, such
as the General Collective Model (GCM) \cite{gcm}, we are not aware of a 
systematic set of GCM calculations, including predictions for the E0 strength, 
for the nuclei in question.

The nature of excited $0^+$ states in $^{154}$Sm is particular
interesting, since $^{154}$Sm is the only rare earth nucleus with
two excited $0^+$ states below the excitation energy of the band
head of the $\gamma$ band at 1440~keV. The excitation energies of
the two $0^+$ states are only 103~keV apart, however, they have very
different properties. As the $0_3^+$ state is only weakly populated in 
Coulomb excitation a small transition strength to the ground-state band
can be concluded \cite{kruecken}. The $0_2^+$
state at 1099~keV has very different properties, the measured lifetime of
1.3(3)~ps results in a rather large transition probability of
$B(\text{E2};\, 0_2^+\rightarrow 2_\text{g}^+)=12(2)$~W.u.. This
leads to the interpretation of the $0_2^+$ state as being the
$\beta$ vibration built on the ground state, while the $0_3^+$ state
cannot be interpreted as a collective excitation and also does not
mix appreciably with the $0_2^+$ state. To confirm this
interpretation, the electric monopole strength $\rho^2(\text{E0};\,
0^+_2 \rightarrow 0^+_\text{g})$ has to be determined.

In $^{166}$Er four excited $0^+$ states are known from
two-neutron transfer experiments \cite{burke}. The $B(\text{E}2)$
values for the transitions from the first three excited $0^+$ states 
to the ground-state band and to the
$\gamma$ band were obtained from lifetime measurements using the
Doppler-shift attenuation method following inelastic neutron
scattering \cite{garrett}. The $0_2^+$ and $0_3^+$ states at 1460~keV
and 1713~keV have small $B(\text{E}2)$ values to both the
ground-state band as well as to the $\gamma$ band. This and the strong
relative population in two-neutron transfer reactions \cite{burke}
suggests  that these states are mainly pair-type excitations.
In contrast, the $0_4^+$ state at 1934~keV has a strong
transition strength branch to the ground-state band ($B(\text{E2};\,
0_4^+\rightarrow 2_\text{g}^+)=8.8 (9)$~W.u.) and no observable
decay to the $\gamma$ band. Thus the $0_4^+$ state is interpreted as
a $\beta$-vibrational state. In addition, a fifth $0^+$ state
was reported in Ref. \cite{garrett1} and interpreted as the $0^+$
member of the $\gamma\gamma$ phonon multiplet, due to its large
$B(\text{E}2;0_5^+ \rightarrow 2_{\gamma}^+)$ value. As mentioned
earlier, the electric monopole strength for the $0_2^+$ state was
measured to $\rho^2(\text{E0};\, 0^+_2 \rightarrow 0^+_\text{g}) = 2.2
(8)\cdot 10^{-3}$ \cite{wood}, hence a rather small value supporting 
the interpretation not to be the $\beta$ vibration.

\section{Setup and experimental procedure}
\label{setup}

Excited states in  $^{154}$Sm and $^{166}$Er were populated via safe
Coulomb excitation using isotopically enriched self-supporting targets (760 and
995~$\mu$g/cm$^2$, respectively) and an $^{16}$O beam from
the Tandem accelerator of the Maier-Leibnitz-Laboratory (MLL) in
Munich ($E_\text{lab}=$55, 60 and 65~MeV). Scattered particles were
detected in a 64-fold segmented double-sided Silicon strip detector
(DSSSD) in backward direction (covering angles from $152^\circ$ to
$170^\circ$). The electrons were registered in a cooled Si(Li)
detector in conjunction with a Mini-Orange (MO) spectrometer.
Simultaneously the $\gamma$ rays emitted by the excited nuclei were
detected with a MINIBALL triple-cluster Germanium detector
\cite{miniball}. A sketch of the setup is shown in Fig.
\ref{setup}.
\begin{figure}[h]
\centering
\includegraphics[width=6.3cm]{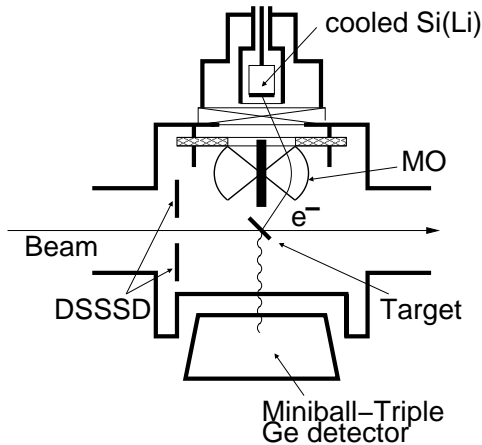}
\caption{Sketch of the experimental setup used at the MLL. Electron
and $\gamma$ detectors were positioned perpendicular to the beam axis. 
Backscattered $^{16}$O nuclei were registered in an
annular DSSSD at backward angles.}\label{setup}
\end{figure}

The Mini-Orange \cite{klinken} consists of 8 wedge-shaped permanent
magnets arranged around a central Pb absorber with a toroidal
field of 160~mT. For the experiment on $^{154}$Sm the
transmission curve of the Mini-Orange was optimized for the expected
E0 transition energy of 1053 keV in resulting in a transmission efficiency of 6.5 \%. For the $^{166}$Er
experiment the maximum of the transmission curve was shifted to
1700 keV in order to measure the expected E0 transitions at 1402,
1656 and 1877 keV simultaneously with transmission efficiencies of 2.5, 3.5 and 2.7 \% respectively. 

The electric monopole strength $\rho^2(\text{E0})$ is used to
characterize E0 transitions. It is given by
\begin{equation}
\rho(\text{E0}; \text{i}\rightarrow \text{f}) = \frac{\left\langle\
\text{f} \left| M(\text{E0}) \right|
\text{i} \right\rangle}{eR^2} \label{eq:rho_def}
\end{equation}
where $R$ is the nuclear radius ($R\simeq 1.2A^{1/3}$~fm) and $M(\text{E0})$ 
is the monopole matrix element. The corresponding
partial lifetime $\tau$(E0) is given by the electric monopole
strength $\rho^2(\rm E0)$ and the non-nuclear electronic factors
$\rm\Omega$:
\begin{equation}
\frac{1}{\tau(\text{E0})} = \rho^2(\text{E0}) \cdot (\Omega_\text{K}
+ \Omega_\text{L} + ~...+ \Omega_\text{IP}) \label{eq:eo_rate}
\end{equation}
Experimentally the monopole strength is determined from the ratio of
E0 and E2 K-conversion intensities $\rm q_K^2$ and the E2 transition
rate $\rm W_\gamma(E2)$ \cite{kibedi}.
\begin{equation}
\rho^2(\text{E0}) = q_\text{K}^2(\text{E0}/\text{E2}) \cdot \frac{\alpha_\text{K}(\text{E2})}{\Omega_\text{K}(\text{E0})} \cdot W_\gamma
= \frac{I_\text{K}(\text{E0})}{I_\text{K}(\text{E2})} \cdot \frac{\alpha_\text{K}(\text{E2})}{\Omega_\text{K}(\text{E0})}\cdot  \frac{1}{\tau_\gamma}
\label{eq:rho_exp}
\end{equation}
The conversion coefficients $\alpha_\text{K}$ and the electronic
factors $\Omega_\text{K}$ are tabulated \cite{kantele}, the lifetime
of the excited $0^+$ states of interest is known from previous
experiments.

\section{Results for $^{154}$Sm}\label{sm}

\begin{figure}[h]
\centering
\includegraphics[width=0.75\textwidth]{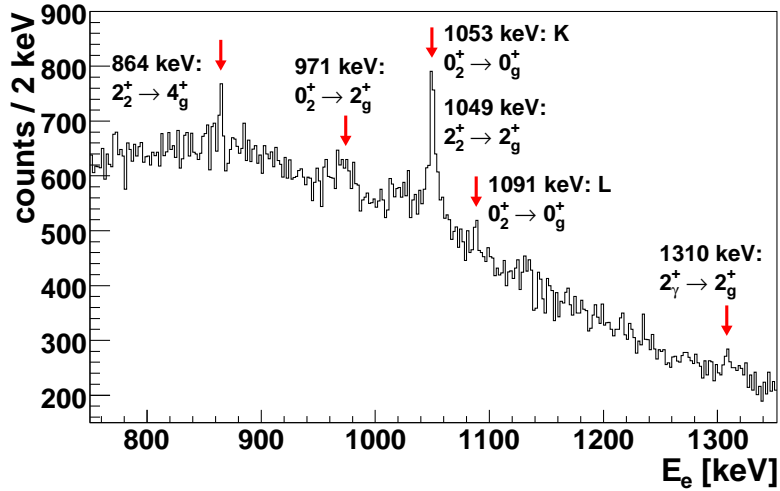}
\caption{Singles energy spectrum of conversion electrons
following Coulomb excitation of $^{154}$Sm.} \label{sm_sili_tot}
\end{figure}

Fig. \ref{sm_sili_tot} shows the $^{154}$Sm conversion electron
singles spectrum for 60~MeV beam energy. 
The $0^+_2\rightarrow 0^+_\text{g}$ and
the $2^+_2\rightarrow 2^+_\text{g}$ transitions in $^{154}$Sm are
only 3.5~keV apart and cannot be separated unambiguously in our
experiment with a detector resolution of 4.6~keV and additional
Doppler broadening. The binding energy for electrons in the K-shell 
amounts to 46.8~keV. Besides the K conversion peak at 1050~keV the 
L conversion can be seen at $E_\text{e}=1091$~keV (binding energy 7.7~keV).

Since the $\rho^2(\text{E0})$ value for the $2^+_2\rightarrow
2^+_\text{g}$ transition is not known, the relative contributions of the two
transitions ($0^+_2\rightarrow 0^+_\text{g}$ and $2^+_2\rightarrow
2^+_\text{g}$) to the peak could not be determined. 
Therefore, the $\rho^2(\text{E0};\, 0^+_2 \rightarrow
0^+_\text{g})$ value could not be deduced from the singles spectrum.
We performed Coulomb excitation calculations showing that the 
excitation probability for multiple excitations
rises with increasing scattering angle. Since the $0^+$ states can
only be excited in multiple-step processes, their excitation
probability rises for large scattering angles, whereas the excitation
probability of the $2_2^+$ state slightly drops with increasing
angle. For particles that are scattered onto the particle detector,
the excitation probability for the $0^+_2$ state is by a factor of
13 larger than for the $2^+_2$ state. Thus for electrons in
coincidence with $^{16}$O ions hitting the DSSSD, the contribution
from the $2^+_2\rightarrow 2^+_\text{g}$ transition can be
neglected, even under the assumption that both E0 transitions have
similar strength.

\begin{figure}[h]
\centering
\includegraphics[width=0.75\textwidth]{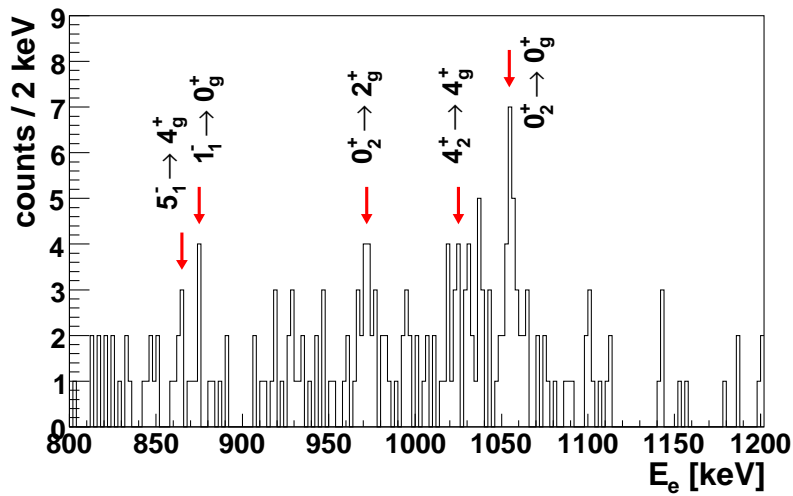}
\caption{Background-subtracted $^{154}$Sm conversion electron
spectrum in coincidence with particles hitting the DSSSD.}
\label{sm_sili_coinc}
\end{figure}

Fig. \ref{sm_sili_coinc} shows the conversion electron
spectrum in coincidence with backscattered projectiles. 
The transitions $0_2^+\rightarrow0_\text{g}^+$ at $E_\text{e}=1053$~keV
and $0_2^+\rightarrow2_\text{g}^+$ at $E_\text{e}=971$~keV from the
first excited $0^+$ state are the strongest lines in the spectrum.
The observed intensity in this spectrum is $23.4 (48)$ counts in the
E0 K-conversion transition line at $E_\text{e}=1053$~keV and $10.4
(32)$ counts in the E2 transition measured in 55~h beam time. The K
conversion coefficient for the 1018~keV E2 transition in $^{154}$Sm
is $\alpha_\text{K}(\text{E2}) = 2.045\cdot10^{-3}$ and the $\Omega$
factor is $\Omega_\text{K}(\text{E0}) = 3.688\cdot 10^{10}$~s$^{-1}$
\cite{kantele}. The lifetime of the first excited $0^+$ state has
been measured to $\tau = 1.3(3)$~ps \cite{kruecken}. Thus a value of
$\rho^2(\text{E0};\, 0^+_2 \rightarrow 0^+_\text{g}) = 96 (42)\cdot
10^{-3}$ can be extracted.

With this value now also the electric monopole strength for the 
$2^+_2 \rightarrow 2^+_1$ transition can also be determined from 
the number of counts in the peak in Fig. \ref{sm_sili_tot} . The ratio
$q_\text{K}^2(\text{E0}/\text{E2})$ for the $2^+_2 \rightarrow
2^+_1$ transition can be determined to be smaller than $0.97$.
The $\Omega$ factor is $\Omega_\text{K}(\text{E0}) = 3.65\cdot
10^{10}$~s$^{-1}$ \cite{kantele} and the lifetime of $2_2^+$ state
has been measured in deuteron scattering ($\tau=2.36(60)$~ps
from $B(\text{E2};\,0_\text{g}^+\rightarrow 2_2^+)=0.020(5)$~$e^2$b$^2$ 
\cite{veje}) and Coulomb excitation ($\tau > 3.5$~ps
\cite{kruecken}). Using $\tau=2.36(60)$~ps (partial lifetime of the
transition $2^+_2 \rightarrow 2^+_1$: $5.8(15)$~ps) one obtains
$\rho^2(\text{E0};\, 2^+_2 \rightarrow 2^+_1)<8.1\cdot 10^{-3}$.
However, this lifetime is not consistent with the present Coulomb
excitation yield, which can only be reproduced with a $B(\text{E2})$ 
value corresponding to a lifetime of $3.0(5)$~ps. This leads to an 
upper limit for the electric monopole strength of $\rho^2(\text{E0};\, 
2^+_2\rightarrow 2^+_1) < 6.3\cdot 10^{-3}$. This is a surprisingly low 
value and we will come back to this in the discussion.

\section{Results for $^{166}$Er}\label{er}

Two separate experiments with the $^{166}$Er target have been performed at
55 and 65~MeV beam energy. The excitation of the $0_4^+$ state is
clearly visible in the $\gamma$-ray energy spectra (not shown)
by the observation of the 1854~keV $0^+_4\rightarrow 2^+_\text{g}$
transition. For the determination of the electric monopole strength
in $^{166}$Er only the conversion electron singles spectra could be
used. The amount of conversion electron and particle detector
coincidences attributed to Coulomb excitation reactions was less than
1 event/keV in 35 and 37.45 h run time respectively, for the two
beam energies.

\begin{figure}[h]
\centering
\includegraphics[width=0.75\textwidth]{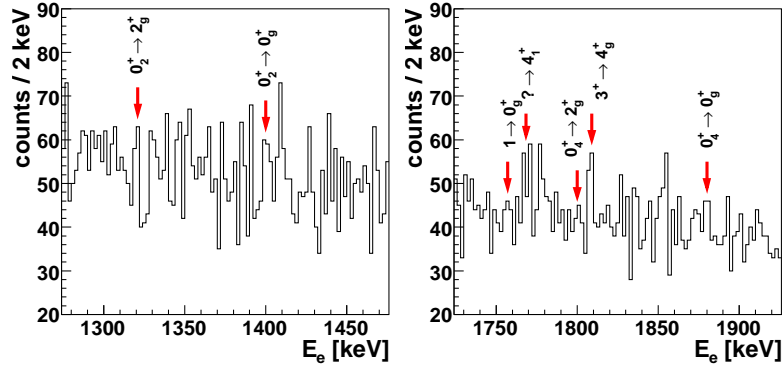}
\caption{Singles spectrum of conversion electrons
following the Coulomb excitation of $^{166}$Er at 55~MeV beam
energy.} \label{er_sili_tot}
\end{figure}

Fig. \ref{er_sili_tot} shows the singles energy spectrum of
conversion electrons for 55~MeV beam energy. The binding energy 
of the K-shell amounts to 57.5~keV. The E0 and E2
transitions from the $0^+_2$ and the $0^+_4$ state could be
identified in the spectrum, despite the poor statistics as 
the energies are known.

The combined statistics of the two experiments at both energies 
allowed to determine for the decay of the $0^+_2$ state an intensity 
ratio of $q_\text{K}^2(\text{E0}/\text{E2}) = 0.47(19)$,
taking into account the ratio of the transmission of the Mini-Orange
for the E0 and the E2 transition. The $\Omega_\text{K}(\text{E0})$
factor for 1460~keV transition energy in $^{166}$Er amounts to
$\Omega_\text{K}(\text{E0})=1.201\cdot 10^{11}\text{s}^{-1}$, the
conversion coefficient for K conversion is
$\alpha_\text{K}(\text{E2}) = 1.505\cdot 10^{-3}$ \cite{kantele}.
The lifetime of the $0_2^+$ state in $^{166}$Er has been measured
to $\tau = 1.1 (4)$~ps \cite{garrett}. For the $0^+_2$ state an
electric monopole strength of $\rho^2(\text{E0};\, 0^+_2 \rightarrow
0^+_\text{g})= 5.3 (23)\cdot 10^{-3}$ (average value of our two
measurements) could be determined which is slightly larger than the
previously known value of $\rho^2(\text{E0};\, 0^+_2 \rightarrow
0^+_\text{g})= 2.2 (8)\cdot 10^{-3}$ \cite{wood} but not strictly 
contradicting.

For the 1934~keV E0 transition from the $0_4^+$ state the electronic factor 
amounts to $\Omega_\text{K}(\text{E0})=1.702\cdot10^{11}\text{s}^{-1}$ and 
$\alpha_\text{K}(\text{E2}) = 8.719\cdot10^{-4}$. The observed intensity 
ratio $q_\text{K}^2(\text{E0}/\text{E2}) = 1.94 (91)$ and the lifetime of
the $0^+_4$ state of $\tau = 78(8)$~fs \cite{garrett} result in a
$\rho^2(\text{E0};\, 0^+_4 \rightarrow 0^+_\text{g})= 127(60)\cdot
10^{-3}$.

\section{Discussion}
\label{discussion}

The E0 measurements on $^{166}$Er and $^{154}$Sm presented here have
revealed, despite the significant experimental uncertainties, large 
$\rho^2(\text{E0})$ values from the states that have
previously been associated with $\beta$-vibrational states in these
well-deformed rare earth nuclei. The results also generally confirm
for the first time the recent predictions by the IBA
model \cite{brentano} of large $\rho^2(\text{E0};\,0^+ \rightarrow
0^+_\text{g})$ values. However, as the following comparison to ECQF IBA
calculations will show, the situation is not quite as
straightforward and a number of open questions will remain.

This confirmation of large E0 strength from $\beta$-vibrational states
is also supported by the results of a re-analysis of published
conversion electron data for $^{240}$Pu. We found that in the
superdeformed second minimum of the potential surface
\cite{gassmann} an average monopole strength of
$\rho^2(\text{E0};\,I_\beta^+\rightarrow I_\text{g}^+) = 55
(24)\cdot 10^{-3}$ could be determined for the $\beta$-vibrational
band members at 785.1~keV ($2_2^+$), 825.0~keV ($4_2^+$), 892.4~keV
($6_2^+$) and 986.8~keV ($8_2^+$).

\subsection*{$^{154}$Sm}

$^{154}$Sm is the only nucleus in the rare earth region with two
excited $0^+$ states below the excitation energy of the band head of the
$\gamma$-vibrational band (1440~keV). Although these two states are
only 103~keV apart, they have very different
properties. The $0^+_2$ state has a strong collective transition to
the ground-state band, $B(\text{E2}\, 0_2^+\rightarrow 2_\text{g}^+)
= 12(2)$~W.u. \cite{kruecken}, while for the $0^+_3$ state only an
upper limit $B(\text{E2}\, 0_3^+\rightarrow
2_\text{g}^+) <0.3$~W.u. is known. This is consistent with the
interpretation that the $0^+_2$ state is the ground state of the
$\beta$-vibrational band. Since $^{154}$Sm is only two neutrons away
from the critical point nucleus $^{152}$Sm, the $0_3^+$ state may
well be the spherical shape-coexisting $0^+$ state.

\begin{figure}[h]
\centering
\includegraphics[width=\textwidth]{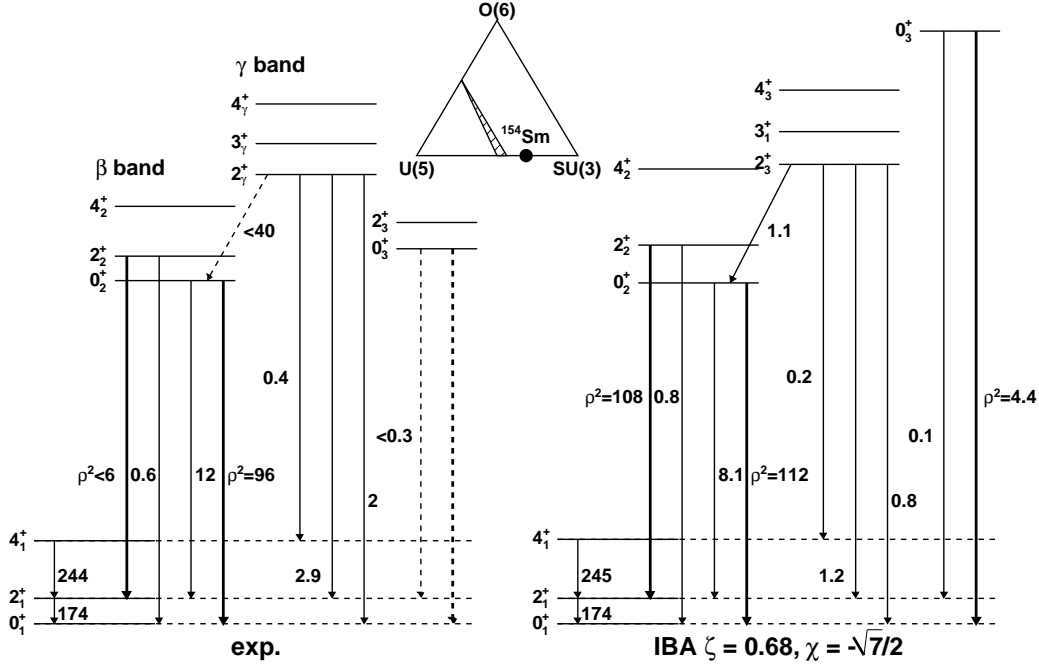}
\caption{$^{154}$Sm level scheme in comparison with IBA
calculations. E0 transitions are marked with thick lines. Dashed lines indicate transitions which have not been observed. The inset shows the position of $^{154}$Sm within the IBA symmetry triangle.}
\label{fig:ibaexpsm}
\end{figure}

Fig. \ref{fig:ibaexpsm} shows the level scheme for the lowest
positive parity states in $^{154}$Sm together with the IBA
prediction for the parameter pair $\chi=-\sqrt{7}/2$ and
$\zeta=0.68$ \cite{scholten, hinke2}, thus positioning
$^{154}$Sm directly on the $U(5)-SU(3)$ leg of the symmetry triangle.
In this region of the IBA the $\gamma$ band is rather high in energy
and the lowest excited $0^+$ state has a collective E2 transition to
the ground-state band and a strong E0 transition to the ground
state. For the IBA calculations the level energies are scaled
to the experimental $E(2^+_1)=82.0$~keV, while the transition rates
(indicated for each transition in Weisskopf units, W.u.) are scaled
to the experimental $B(\text{E2};\, 2^+_1\rightarrow
0^+_\text{g})=174$~W.u..

The experimental properties of the $0_2^+$ state are well described by the 
IBA calculations. For the $0^+_2$ state the large measured monopole 
strength of $\rho^2(\text{E0};\, 0^+_2 \rightarrow 0^+_\text{g}) = 
96(42)\cdot 10^{-3}$, measured within this work, and the collective E2 
decay to the ground state band confirm the interpretation of the $0^+_2$ 
state as $\beta$-vibrational state. The experimental fact that the $0_3^+$ 
state is non-collective is reproduced in the IBA calculations which also 
show it to have a dominant contribution for the number of $d$ bosons $n_d=0$, 
consistent with the interpretation as spherical shape-coexisting state 
(see Fig. 3 in Ref. \cite{iachello}). 

While the situation for the $0^+$ states seems satisfactory, it is however 
not for the $2_2^+$ state, which is considered to be the $2^+$ member of 
the $\beta$-vibrational band. This state exhibits a transition strength 
to the  ground state of $B(\text{E2};\,2^+_2\rightarrow 0^+_\text{g})=0.64(12)$~W.u. consistent with the 0.76 W.u. predicted by the IBA, but at the same time 
a surprisingly small electric monopole strength of 
$\rho^2(\text{E0};\, 2^+_2 \rightarrow 2^+_\text{g}) < 6.3\cdot10^{-3}$ was 
measured, which is almost 16 times smaller than the expectation that
the E0 strength of the $2^+_2 \rightarrow 2^+_\text{g}$ transition should
be about the same as that of the $0^+_2 \rightarrow 0^+_\text{g}$ monopole
transition.

A possible explanation for this behavior may lie in a mixing of
the $2_2^+$ state with other $2^+$ states. The obvious candidate
would be the $2_3^+$ state at 1286~keV which belongs to the
$K^\pi=0^+$ band built upon the $0_3^+$ state. However, the $0_2^+$
and $0_3^+$ states exhibit at most a $4\%$ mixing \cite{kruecken},
making it unlikely that the mixing of the $2_2^+$ and $2_3^+$
should be significantly larger. It may be more likely that there is
significant mixing of the $\beta$ and $\gamma$ bands. However, the 
calculation of mixing amplitudes does not lead to consistent values, 
the branching ratios cannot be explained by a simple band 
mixing model. The mixing amplitude cannot be extracted quantitatively, 
however, the calculations reveal only a small mixing between 
$\beta$ band, $\gamma$ band and ground-state band. 
However, as shown in \cite{brentano}, in the IBA
the total E0 strength depends on the subtle sum of many $n_d$
components, some with positive and some with negative sign. Thus
even small admixtures of other states may lead to subtle but
decisive changes of the $n_d$ distribution, possibly leading to the
cancellation of the E0 strength.

\subsection*{$^{166}$Er}

\begin{figure}[h]
\centering
\includegraphics[width=\textwidth]{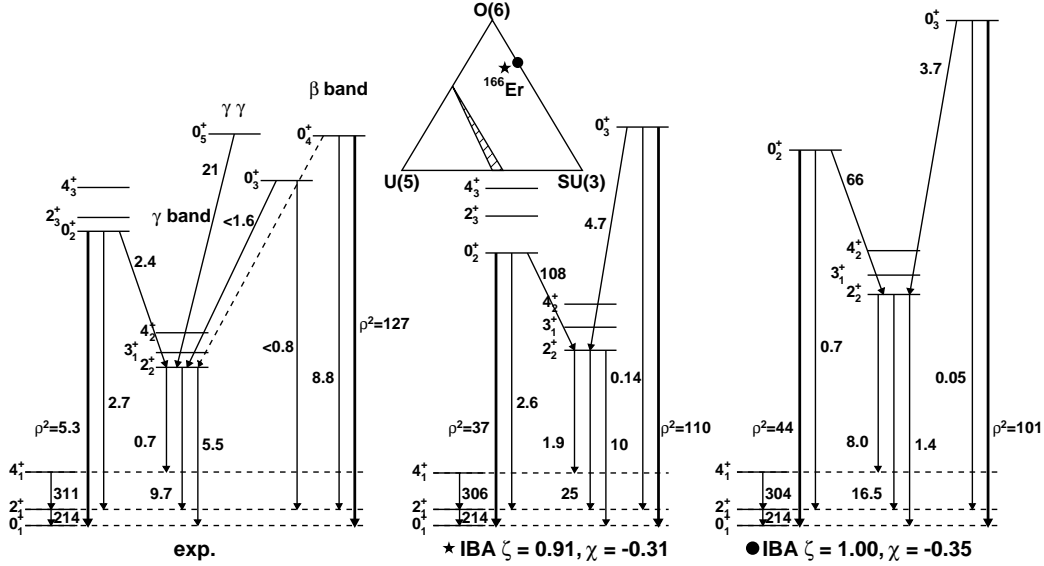}
\caption{$^{166}$Er level scheme in comparison with IBA
calculations for the parameter pairs $(\zeta, \chi)=(0.91,-0.31)$ \cite{mccutchan} and $(\zeta, \chi)=(1.0,-0.35)$ in order to account for the high excitation energy of the $0_{\gamma\gamma}^+$ state. E0 transitions are marked with thick lines. Dashed lines indicate transitions which have not been observed. The inset indicates the position of the two IBA parameter sets within the IBA symmetry triangle.}
\label{fig:ibaexper}
\end{figure}

Fig. \ref{fig:ibaexper} shows part of the $^{166}$Er level scheme
with its five $0^+$ states below 2 MeV in comparison with IBA
calculations with the parameters obtained in Ref. \cite{mccutchan},
placing it near the O(6) corner of the symmetry triangle, although
still being well-deformed with no significant $\gamma$ softness, as
attested by the $R_{4/2}=E(4_1^+)/E(2_1^+)$ ratio of 3.29. For the IBA calculations the level energies are scaled to the experimental $E(2^+_1)=80.6$~keV, while the transition rates (indicated for each transition in Weisskopf units, W.u.) are scaled to the experimental $B(\text{E2};\, 2^+_1\rightarrow 0^+_\text{g})=214$~W.u..

The $0_2^+$ state has no collective E2 transition to the
ground-state band ($B(\text{E2};\,0_2^+\rightarrow 2_\text{g}^+) =
2.7 (10)$~W.u.) nor to the $\gamma$ band ($B(\text{E2};\,
0_2^+\rightarrow 2_\gamma^+) = 2.4(7)$~W.u.) \cite{garrett}
and only a small $\rho^2(\text{E0})$ to the ground state, which was
confirmed in this experiment. For the $0_3^+$ state only upper
limits for its decay to the ground-state band and the $\gamma$ band
are known. This state has not been excited in this Coulomb
excitation experiment and no E0 strength is known, but it is clear
that this state is not a collective excitation of the ground state.
Due to their non-collective behavior and the rather strong
excitation via two-neutron transfer, both the $0_3^+$ state and the
$0_2^+$ state have been interpreted as dominated by pair excitations
\cite{garrett}. Thus they are beyond the scope of the framework of
the IBA. However, the energy of the $0_2^+$ state has been used in
the fits of Ref. \cite{mccutchan} to determine the IBA parameters,
explicitly assuming that this state is collective in nature.

However, the IBA parameters do not change dramatically, if
the $0_3^+$, $0_4^+$ or even $0_5^+$ states are used, as
is apparent from Fig. 3 of Ref. \cite{mccutchan1} because the ratio
$R_{0\gamma}=[E(0_2^+)-E(2_{\gamma}^+)]/E(2_1^+)$ changes
from 8.3 to 14.2 and $^{166}$Er is located slightly closer to
the $O(6)-SU(3)$ leg of the symmetry triangle. The excitation energy of the $0_{\gamma\gamma}^+$ state at 1943~keV can be approximately reproduced in the ECQF using IBA parameters $(\zeta, \chi)=(1.0,-0.35)$. 

In this region of the IBA symmetry triangle, the $0_2^+$ state shows
a very collective decay to the $2_{\gamma}^+$ state and only weak
transitions to the ground state band, being consistent with the
interpretation of this state being the $0^+$ member of the
$\gamma\gamma$-phonon multiplet. These decay properties are most
consistent with that of the experimental $0_5^+$
state \cite{garrett1}. At the same time the $0_3^+$ state in the IBA
calculations shows a large E0 strength to the ground state, which
would be consistent with the expectation for a $\beta$-vibrational
state. However, for this region of the IBA symmetry triangle, no
excited $0^+$ state exhibits a collective E2 decay to the ground
state band, which would be a prerequisite for this interpretation.
Thus it seems that in the IBA there exists no state in this parameter
range that is consistent with the traditional concept of a $\beta$
vibration (namely large $\rho^2(\text{E0})$ and large 
$B(\text{E2};\,0^+ \rightarrow 2_\text{g}^+)$).

However, the experimental situation in $^{166}$Er is not consistent
with this IBA picture, since the $0_4^+$ state does exhibit a strong
collective transition to the ground-state band ($B(\text{E2}\,
0_4^+\rightarrow 2_\text{g}^+) = 8.8 (9)$~W.u. \cite{garrett}) and
no observable transition to the $\gamma$ band. Moreover, two-nucleon
transfer reactions showed that the (p,t) cross section is low, which
led to the interpretation that the $0_4^+$ state is the band head of
the $\beta$-vibrational band \cite{garrett}. The large value of 
$\rho^2(\text{E0})= 127(60)\cdot 10^{-3}$ to the ground state 
obtained in this work is consistent with this interpretation.

Thus, even by considering the experimental $0_2^+$ and $0_3^+$ state
as non-collective and therefore not within the framework of the IBA the transition properties of the $0^+$ states cannot be reproduced.

\section{Conclusion}
\label{concl}

Excited states in the well-deformed rare earth isotopes $^{166}$Er
and  $^{154}$Sm were populated via Coulomb excitation at the MLL
Tandem accelerator. Conversion electrons were registered in a cooled
Si(Li) detector in conjunction with a magnetic transport and filter
system, the Mini-Orange.

For the first excited $0^+$ state in $^{154}$Sm at 1099~keV a
large value of the monopole strength for the transition to the
ground state of $\rho^2(\text{E0};\, 0^+_2 \rightarrow 0^+_\text{g})
= 96 (42)\cdot 10^{-3}$ was extracted. This confirms the
interpretation of the lowest excited $0^+$ state in $^{154}$Sm as
the collective $\beta$-vibrational excitation of the ground state.

In $^{166}$Er we observed E0 transitions from the $0^+_2$ as well as from the
$0^+_4$ state. For the $0^+_2$ state we obtained a
value of $\rho^2(\text{E0};\, 0^+_2 \rightarrow 0^+_\text{g}) = 5.3
(23)\cdot 10^{-3}$ in agreement with the known value of $ 2.2
(8)\cdot10^{-3}$ \cite{wood}. The newly measured large electric
monopole strength of $\rho^2(\text{E0};\, 0^+_4 \rightarrow 0^+_\text{g}) =
127 (60)\cdot 10^{-3}$ is consistent with the previous assignment
\cite{garrett} of the $0_4^+$ state at 1934~keV to be the
$\beta$-vibrational excitation of the ground state.

In a re-analysis of published conversion electron data for
$^{240}$Pu in the superdeformed second minimum of the potential
surface \cite{gassmann} an average monopole strength of
$\rho^2(\text{E0};\,I_\beta^+\rightarrow I_\text{g}^+) = 55
(24)\cdot 10^{-3}$ could be determined for the $\beta$-vibrational
band members up to the $8_2^+$ state.

The observed large monopole strength in all three deformed nuclei
for the first time experimentally confirms the theoretical
predictions \cite{brentano} that the lowest excited $0^+$ states in
deformed nuclei exhibit strong monopole transitions to the ground
state.

A more detailed comparison of the level schemes of the two
rare earth nuclei with ECQF IBA calculations reveals that not all
experimental features are reproduced by the IBA. In the region of
the IBA symmetry triangle where the $\gamma$-vibrational band is at
relatively low energy and the first excited $0^+$ state is well
above the $2_{\gamma}^+$ state no excited $0^+$ state shows
collective E2 strength to the ground state band while the $0_2^+$ or
$0_3^+$ states have large E0 strength to the ground state. In this
region of IBA parameters the $0_2^+$ state has the characteristics of a
double-$\gamma$ vibration but no $0^+$ state with the characteristics
of a traditional $\beta$ vibration exists. The case of $^{166}$Er,
where a $\beta$-vibrational state has been clearly observed, seems to
be in contradiction to that feature of the IBA calculations. The
appearance of the low-lying non-collective $0_2^+$ and $0_3^+$ states 
and the fact that the $0_5^+$ double-$\gamma$ vibrational state is 
almost degenerate with the $\beta$-vibrational $0^+_4$ state make this a 
quite unusual case.

Near the $U(5)-SU(3)$ leg of the IBA symmetry triangle the $0_2^+$
state in deformed nuclei lies below the $2_{\gamma}^+$ state and
exhibits all characteristics of a $\beta$-vibrational excitation.
$^{154}$Sm seems to be a very good example of this situation.
However, the properties of the $2_2^+$ state are not in agreement with
the IBA predictions probably due to a mixing with other
$2^+$ states.

We conclude that the two nuclei $^{154}$Sm and $^{166}$Er are 
in general representative for two regions in the IBA triangle, one 
with low lying $\beta$ vibration near the $U(5)-SU(3)$ leg and one 
closer to the $O(6)$ corner (but still with $R_{4/2}\geq 3.1$) with 
the $0_2^+$ state being the two phonon $\gamma\gamma$ vibration but 
without a $0^+$ 
state with the characteristics of a $\beta$ vibration (namely large 
$\rho^2(\text{E0})$ and large $B(\text{E2};\,0^+ \rightarrow 2_\text{g}^+)$). 
However, significant discrepancies in some details are observed and it is an interesting question if other collective models, such as the GCM, may be able to obtain better agreement with the experimental data. However, no systematic studies exist at this point.

From the current investigation, we draw the conclusion, that it is very
important to obtain as much detailed experimental information on all
low-lying $0^+$ states as possible, including data on transfer strength as well
as electromagnetic decay properties. In many cases, where only
partial information on excited $0^+$ states is available, it is not
clear that these states are indeed the ones described in the
framework of the collective models. In addition, mixing of different
structures can lead to significant modifications of the properties,
leading to large deviations from the simple expectations, which
should, if at all, just be used as guiding principles.

{\bf Acknowledgement}\\
We acknowledge fruitful discussions with Prof. K. Heyde, N.V.
Zamfir and R.F. Casten. This work was supported by the DFG Cluster of Excellence ``Origin and Structure of the Universe''.

\end{document}